%% file: topo.tex
\documentclass[11pt,a4paper,twoside]{article}
\usepackage{amsmath,amssymb,amsthm}
\usepackage{graphicx}

\newcommand{\ccc}{{\mathbb C}}

\newcommand{\cc}{{\cal C}}
\newcommand{\dd}{{\cal D}}

\renewcommand{\ss}{{\cal S}}

\newcommand{\zz}{{\cal Z}}

\renewcommand{\t}{\mathrm{Tr}}

\newcommand{\bbb}{\begin{equation}}
\newcommand{\eee}{\end{equation}}
\newcommand{\bbbb}{\begin{eqnarray}}
\newcommand{\eeee}{\end{eqnarray}}

\DeclareMathOperator{\id}{id}
\newcommand{\one}{\mathbf{1}}
\newcommand{\tens}{\otimes}
\newcommand{\defeq}{:=}
 
\newtheorem{thm}{Theorem}[section]
\newtheorem{lem}{Lemma}[section]

\newtheorem{pro}{Proposition}[section]

\allowdisplaybreaks

\begin{document}

\input{titlepage}
\input{intro}
\input{bfdia}

\input{qgrp}
\input{topinv}
\input{geninv}
\input{conclude}
\input{acknowledge}

\bibliographystyle{amsordx}
\bibliography{stdrefs,ref}
\end{document}

%% file: titlepage.tex
\begin{titlepage}
\setcounter{footnote}{0}
\renewcommand{\thefootnote}{\arabic{footnote}}

\title{\textbf{Spin Foam Diagrammatics and \\ Topological Invariance}}

\vskip 1.5cm

\author{Florian
GIRELLI\footnote{\texttt{girelli@cpt.univ-mrs.fr}}\,${}^{a}$${}^{b}$ \
, Robert OECKL\footnote{
\texttt{oeckl@cpt.univ-mrs.fr}}
\,${}^{a}$ \ , Alejandro
PEREZ\footnote{\texttt{perez@gravity.phys.psu.edu}}
\,${}^{a}$${}^{c}$\\[.3cm]
${}^a$ Centre de Physique Th\'eorique, Marseille, France\\[.1cm]
${}^b$ Universit\'e de Provence (Aix-Marseille I), Marseille, France\\[.1cm]
${}^c$ Center for Gravitational Physics and Geometry,\\
Pennsylvania State University, USA} 
\date{CPT-2001/P.4262, CGPG-01/10-1\\ 7 November 2001} 

\maketitle

\vskip 2cm
\centerline{\bf Abstract}

\medskip

We provide a simple proof of the topological invariance of
the Turaev-Viro model (corresponding to simplicial 3d pure Euclidean
gravity with cosmological constant) by means of a novel diagrammatic
formulation of
the state sum models for quantum BF-theories. Moreover, we prove
the invariance under more general conditions allowing the state sum to
be defined
on arbitrary cellular decompositions of the underlying
manifold. Invariance is governed by
a set of identities corresponding to local gluing and rearrangement of
cells in the
complex. Due to the fully algebraic nature of these identities our
results extend to
a vast class of quantum groups. 
The techniques introduced here could be relevant for investigating the
scaling properties of non-topological state sums, being proposed as 
models of quantum gravity in 4d, under refinement of the 
cellular decomposition.

\vskip 1truecm

\vskip 1truecm

\end{titlepage}

%% file: intro.tex
\section*{Introduction}

Since the 1980's, topology of low dimensional manifolds is at the
meeting point of many different areas of mathematics and
physics. This development was initiated by Donaldson
who was studying 4-manifolds by the means of Yang-Mills
equations. Then Jones found his famous polynomial of links in the
3-sphere, using algebraic structures: Von Neumann algebras. This
opened up a new way for algebra to come into low dimensional
topology. Quantum groups play an important role through the
Reshetikhin-Turaev functor
\cite{ReTu:ribboninv}. This latter gives a diagrammatic approach to
categories, and enables one to construct new topological invariants of
3-manifolds. A state sum invariant of this type was first given by
Turaev and Viro
\cite{TuVi:inv3} using an arbitrary
triangulation of the manifold. This state sum is expressed in terms of
$6j$-symbols of $U_q(\mathfrak{sl}_2)$, which are associated with tetrahedra of
the triangulation.
Invariance of the state sum under change of triangulation
implies topological invariance.
The proof of invariance in \cite{TuVi:inv3}
relied heavily on properties of the $U_q(\mathfrak{sl}_2)$ $6j$-symbols
(Biedenharn-Elliot property) and so was mainly algebraic.
Later, the invariant was generalized to arbitrary spherical categories
\cite{BaWe:invplm}, again using algebraic properties of (now
generalized) $6j$-symbols.

State sum invariants arise in the quantization of BF-theory
due to the absence  of local degrees of freedom. 
A typical  example is the case of 3d Euclidean gravity with 
cosmological constant represented by the Turaev-Viro 
model (see \cite{lectures} and references therein). 
Models of quantum gravity in four dimensions have been defined 
by means of appropriately constraining the state sum models for 4d 
BF-theory \cite{BaCr:relsnet, BFtoGR1, BFtoGR2, BFtoGR3}. 
Constraining the state sum  restores the  
local degrees of freedom of gravity while topological invariance is lost.
Physical amplitudes are expected to be recovered by means of summing over 
discretizations or by an appropriate limiting procedure in which the 
discretization is refined. We expect that the techniques 
introduced here would shed light on the issues involved in 
studying the scaling properties of amplitudes in non-topological models
as well. 

A new diagrammatic approach to spin foam models
has been introduced in \cite{Oe:qlgt} where generalized lattice gauge
theory is considered. 
The aim of this article is to
show that this allows for a very simple proof of topological
invariance of 3d BF-theory (which is a special case of lattice gauge
theory).  This applies to the group case
(Ponzano-Regge \cite{PoRe:limracah}), the supersymmetric case, and the
quantum group case (Turaev-Viro \cite{TuVi:inv3}, Barrett-Westbury
\cite{BaWe:invplm}). In the group case the invariance of the partition
function can be
rather easily demonstrated through manipulations with delta functions.
These algebraic operations have simple diagrammatic analogues (which
are even more intuitive than the handling of delta functions).
If one considers
the 3d BF-amplitude with (any) quantum groups, the traditional proof
of invariance (through algebraic manipulations) is much less trivial,
and relies heavily on properties of generalized $6j$-symbols
\cite{Tur:qinv,BaWe:invplm}. We show here that the diagrammatics defined
for ordinary groups extends naturally to the quantum groups case and
therefore renders the proof of topological invariance just as easy.
Instead of complicated algebraic relations and dealing with
$6j$-symbols, the proof requires only a few pictures.

What is more, the new diagrammatic approach allows the
definition of the partition function on arbitrary cellular
decompositions (as opposed to just triangulations) without any
additional complications.
Indeed, contrary to what one might expect, this renders the proof of
invariance even simpler.
While first performing the proof in the simplicial case employing
Pachner moves \cite{Pac:plmfd} we then refine
it to the general cellular case. To this end we introduce a set of
moves relating cellular decompositions. As it turns out, these
correspond to diagrammatic identities of the partition function which
are in a sense elementary, rendering the invariance proof particularly
simple.

In Section~\ref{sec:dia}, we introduce discretized 3d BF-theory and
introduce the diagrammatic language for expressing the partition
function. In Section~\ref{sec:qg}, we discuss the generalization to the
quantum group case. In Section~\ref{sec:topinv}, we
give the proof of topological invariance first for the simplicial case
and extend it then to the general cellular case. To this end we
introduce moves relating a cellular decomposition to a simplicial
one. We end with concluding remarks.
Among other things we discuss there how the presented moves are
extended to relate arbitrary cellular decompositions.

%% file: bfdia.tex
\section{Diagrammatics for BF-theory}
\label{sec:dia}

In this section we introduce BF-theory,
its quantization on a discretized manifold,
and a diagrammatic formalism to represent its partition function
similar to the spin foam/network formalism.

\subsection{The partition function}

Let  $M$ be an oriented compact piecewise-linear manifold, $G$  a
compact Lie group and $P$ a  principal $G$-bundle over
$M$.\footnote{Note that we do note require the bundle to be
trivial. Indeed any principal G-bundle will lead to the same
discretized partition function as any transition function are
``integrated out''.} Let $A$ be a
connection on $P$, and $B$ an ad($P$)-valued 1-form on
$M$. Here, ad($P$) is the associated vector bundle to $P$ via the
adjoint action of $G$ on its Lie algebra. Let $F$ be the curvature
2-form of $A$. The action is then defined as
\begin{equation}
\ss = \int _M \t( B \wedge F) ,
\end{equation}
where $\t$ is the trace in the adjoint representation.
The partition function of the quantum field theory is formally defined
to be
\begin{equation}\label{Z}
{\cal Z}({\cal M})=\int {\cal D}B \ {\cal D}A \ e^{i \int {\rm Tr}\left[B 
\wedge F\right]} .
\end{equation}
Integrating out the $B$-field one formally obtains
\begin{equation}
\zz = \int \dd A \; \delta (F).
\end{equation}
This last expression can be given a precise meaning if
one replaces $M$ by a cellular decomposition $\cal K$.
We recall the definition of cellular decomposition (a decomposition as
a finite CW-complex, see \cite{Mas:algtop}). The basic element
is a ``cell''. An n-cell is an open ball of dimension $n$. A cellular
decomposition $\mathcal{K}$ of $M$ is a presentation of $M$ as a
disjoint union of cells.

Consider now the complex ${\cal K}^{*}$ dual to ${\cal K}$. That is, the
complex where every $n$-cell is replaced by a corresponding $(\dim M
- n)$-cell. If in $\mathcal{K}$ an $n$-cell is in the boundary of an
$m$-cell then in $\mathcal{K}^*$ this relationship is reversed for the
corresponding cells.
As in lattice gauge theory, one assigns a group element, $g_e$, 
to each
1-cell (denoted edge and labeled by $e$) in $\mathcal{K}^*$,
representing the holonomies of the connection.
Then one
defines the measure over the
discretized connections to be $\prod_e dg_e$, where
$dg$ is the \emph{Haar measure} on $G$. The curvature $F$ is represented by 
the holonomy along edges bounding
2-cells (which are denoted \emph{faces} and labelled by $f$), i.e., the
product $g_{e_1} \cdots g_{e_n}$, where $e_1 \cdots e_n$ denote the
edges bounding the corresponding face. We assume that each face is given an 
arbitrary but fixed orientation.
In this way the partition function is defined as
\begin{equation} \label{R}
{\cal Z}= \int \prod_e dg_e \prod_{\rm f} \delta(g_{e_1} \cdots g_{e_n}).
\end{equation}

Let $\cc _{alg} (G)$ be the  algebra of  complex valued
representative functions on $G$, that is, the functions arising as
matrix elements of finite-dimensional complex representations of
$G$. The Peter-Weyl decomposition asserts then that
\[
\cc _{alg} (G) = \bigoplus_{\rho} (V_{\rho}^*\tens V_{\rho}) ,
\]
where $\rho$ are the finite-dimensional irreducible representations of
$G$ and $V_\rho$ the representation spaces.
Concretely, any function $f$ over the group  $G$ can be written as
\[
f(g)= \sum_{\rho,n,n'} C_{\rho}^{nn'}\rho_{nn'}(g) \; \; \; \forall g\in G
\]
with  coefficients  $ C_{\rho}^{nn'}$. In particular,  the delta function is
\[
\delta (g)= \sum_{\rho} {\rm dim} \rho \, \t(\rho(g)) ,
\]
where the sum runs over irreducible representations as above.
Thus, we can rewrite the partition function as
\begin{gather}
{\cal Z}=\sum \limits_{ {\cal C}:\{{\rm f}\} \rightarrow \{\rho_{\rm
f}\} } 
\ (\prod_{{\rm f}}
{\rm dim}\rho_{\rm f})\  \mathcal{Z}_\mathcal{C}
  \label{Z1}\\
\text{with}\quad
\mathcal{Z}_\mathcal{C} =
\int \ (\prod_{{\rm e}} dg_{\rm e})
  \ \prod_{\mathrm{f}}
{\rm Tr}\left[\rho_{\rm f}(g_{e_1}\cdots g_{e_n})\right] ,
\end{gather}
where $e_1,\dots,e_n$ are the edges bounding the face $f$.
$\cc $ is the set of functions associating to each face $f$ an irreducible 
representation $\rho_{\rm f}$ of $G$.
In order to proceed to the integration, one can now expand the trace
into matrix elements
\[
\t(\rho_{\rm f}(g_{e_1} \hdots g_{e_n} ))= \sum_{p_1, \cdots, p_n}
\rho_{{\rm f}\, p_{1} p_{2}}(g_{e_1} )\hdots
\rho_{{\rm f}\,p_{n}p_{1}}(g_{e_n} )
\]
and collect for each edge the functions taking as value the
corresponding group element.
However, instead of proceeding formally we are going to obtain a
diagrammatic representation of the partition function, or more precisely
of $\mathcal{Z}_\mathcal{C}$.

Note that by construction,  we are using in (\ref{Z1}) the Haar
measure on $G$. On the algebraic functions this is the map
$\int: \cc _{alg} (G) \to\ccc$ given by the projection
\[
\bigoplus_{\rho} V^*_{\rho}\tens V_{\rho} \to V_\one^*\tens
V_\one\cong\ccc,
\]
where $\one$ denotes the trivial representation. Now, define a family
of maps $\{T_{\rho}\}$ for any representation $\rho$ as
\[
T_{\rho}: V_{\rho}\to V_{\rho}\quad\textrm{with}\quad v\mapsto   \int dg \; 
\rho( g) v .
\]
This is simply the projection onto the trivial subrepresentation.
\begin{pro}[\cite{Oe:qlgt}]
\label{prop:T}
$T_{\rho}$ (with $\rho(g): V_\rho\rightarrow V_\rho$) defines a family
of intertwiners with the following properties:
\begin{itemize}
\item[(a)] $T_\one = \id_\one$.
\item[(b)] $T_{\rho}=0$ for $\rho$  irreducible  and non-trivial.
\item[(c)] $T$ is a projector, i.e.\ $T_{\rho}^2=T_{\rho}$.
\item[(d)] $T$ commutes with any intertwiner, i.e.\
for $\Phi:V_{\rho} \to V_{\rho '} $ an intertwiner we have
$T_{\rho'}\circ\Phi=\Phi\circ T_{\rho}$.
\item[(e)] $T$ is self-dual, i.e.\ $(T_{\rho})^*=T_{{\rho}^*}$.
\item[(f)] $T_{\rho}\otimes T_{\rho'}=T_{\rho \otimes \rho'}\circ 
(T_{\rho}\otimes \id_{\rho'})$.
\end{itemize}
\end{pro}

We depict representations by lines and $T_{\rho}$ by an unlabelled
box. E.g.\ in the case of $T$ applied to a tensor product of three
representations
$\rho=\rho_1\tens\rho_2\tens\rho_3$
the diagram is
\begin{equation}
\begin{array}{c}
\includegraphics[width=2cm]{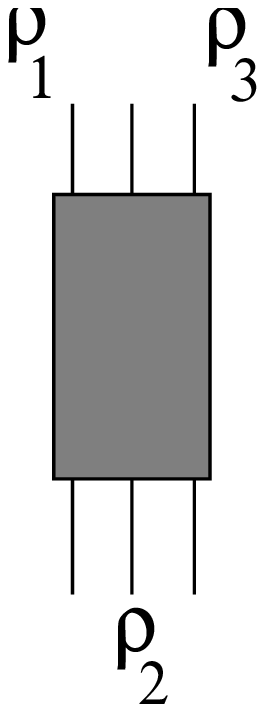}
\end{array}
\begin{array}{c} v_1\tens v_2\tens v_3\mapsto
\int dg \; ( \rho_1(g) v_1 \otimes  \rho_2(g) v_2 \otimes  \rho_3(g) v_3).
\end{array}
\label{put}
\end{equation}

Furthermore, every line carries an arrow and reversal of the arrow 
corresponds to
exchanging the representation associated with that line with its dual.

The diagrammatic representation of the partition function is now
obtained starting from the dual cellular complex ${\cal K}^{*}$ as follows.
Put a loop (called a \emph{wire}) into each
face close to its boundary. Give each wire an arrow according to an
arbitrarily chosen but fixed orientation
the face and the representation label of the face (given by $\mathcal{C}$ in 
Eq.~(\ref{Z1})). Each group integration,
$dg_e$, over the group element associated with the corresponding edge $e$ 
gives rise to
an intertwiner $T_{\rho_1 \otimes \cdots \otimes \rho_N}$ defined above.
${\rho_1 \otimes \cdots \otimes \rho_N}$ is the tensor product of the 
irreducible representations labeling
the faces adjacent to the edge $e$, i.e., the corresponding wires adjacent 
to $e$.
According to Eq.~(\ref{put}) we represent $T_{\rho_1 \otimes \cdots \otimes 
\rho_N}$
by a box or cylinder, denoted \emph{cable}, which we put around the edge.
The terminology comes from the fact that the diagram obtained in this
way resembles an arrangement of cables and wires.

\begin{figure}[h]
\centerline{\hspace{0.5cm}
\( \begin{array}{c}\includegraphics[width=4.0cm]{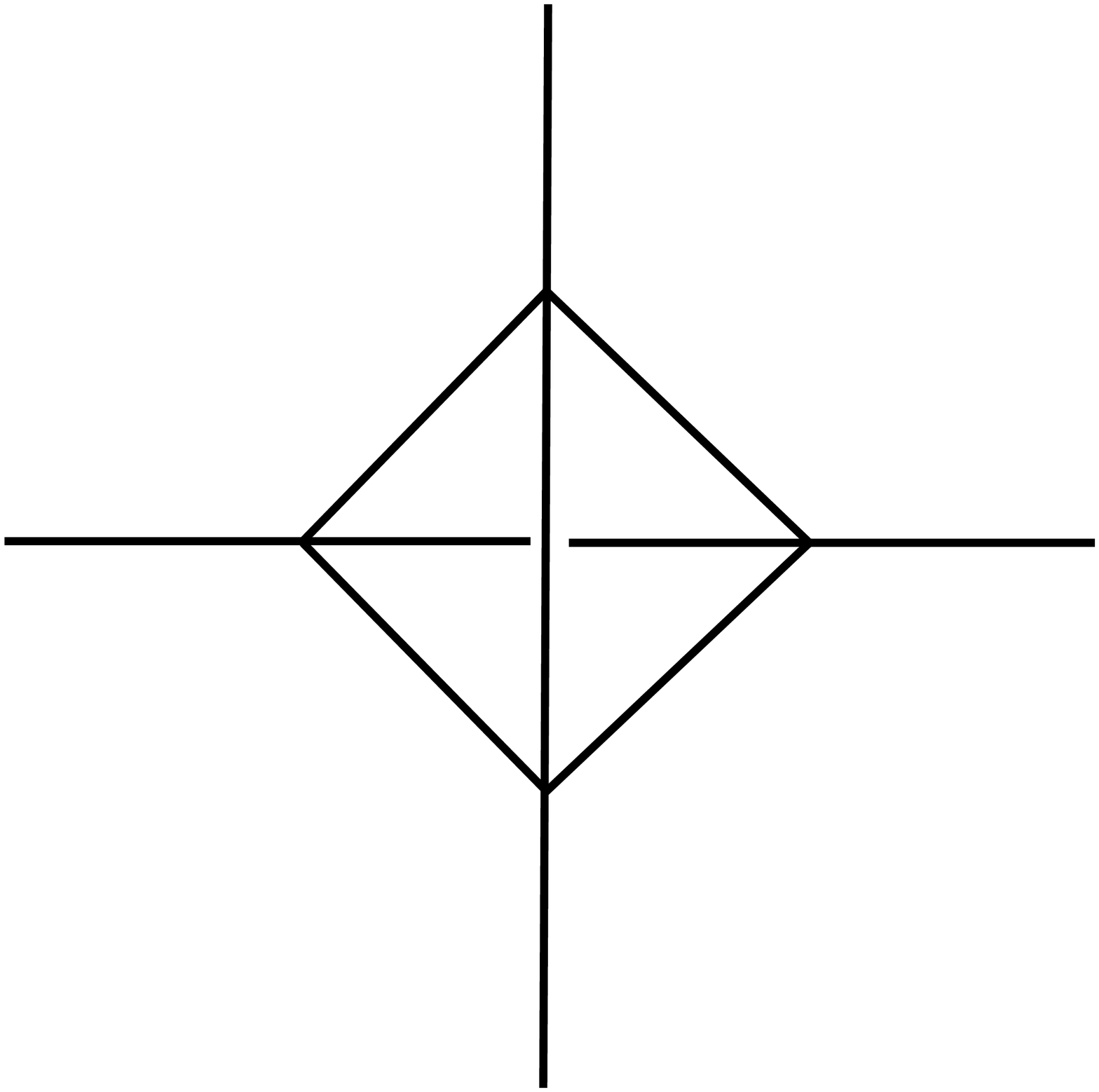}\end{array} 
{\bf  \rightarrow}
\begin{array}{c}\includegraphics[width=4cm]{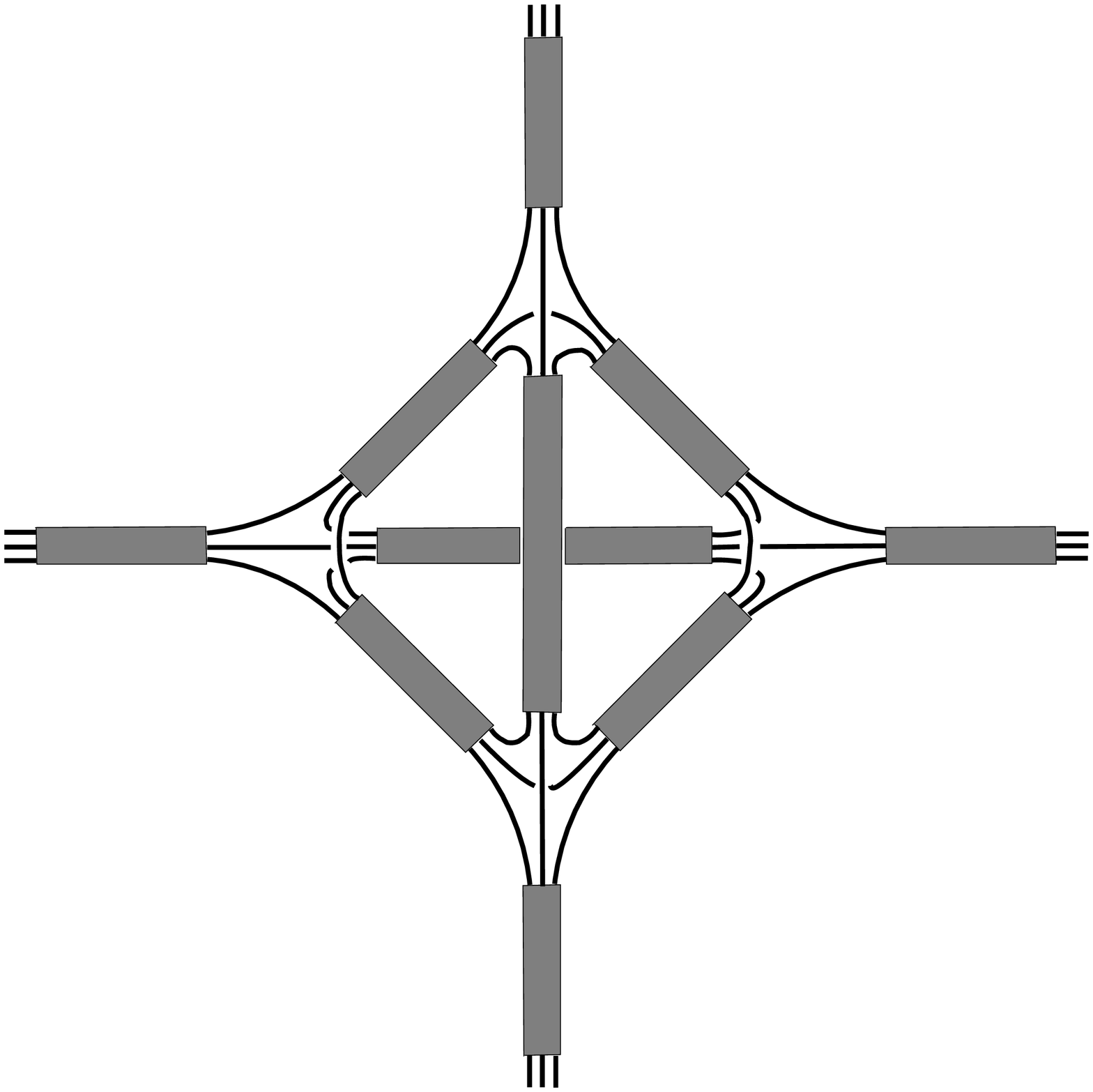} \end{array}\) 
}
\caption{The circuit diagram construction for a
tetrahedron.}
\label{fig:cw}
\end{figure}

In Fig.~{\ref{fig:cw}} we give an illustration of such a diagram. The 
tetrahedron on the
left hand side of Fig.~{\ref{fig:cw}} is part of a bigger cellular
complex $\mathcal{K}^{*}$.
On the right hand side there are six internal cables (cylinders)
corresponding to the six
edges forming the tetrahedron plus four external cables corresponding to 
external edges
that connect the tetrahedron with the rest of $\mathcal{K}^*$. There
are four internal faces, the four triangles
bounding the tetrahedron, represented by closed wires going from one cable 
to the other.

We call the whole (embedded) diagram the \emph{circuit diagram}.
To extract the value of the partition function (or more
precisely $\mathcal{Z}_\mathcal{C}$) one projects the
circuit diagram (arbitrarily) onto the plane. The evaluation proceeds
by using (the symmetric category version of) the Reshetikin-Turaev
functor, see \cite{Oe:qlgt} for details.
Technically, every element of the (projected) circuit diagram represents a
morphism (intertwiner) in the category of representations of $G$. In
particular, the circuit diagram as a whole labelled by the collection
$\mathcal{C}$ represents an intertwiner $\ccc\to\ccc$ and thus a complex
number which is precisely $\mathcal{Z}_\mathcal{C}$.
This is crucial for the generalization to the quantum group case (see
Section~\ref{sec:qg}).

Note that this construction generalizes from
BF-theory to
lattice gauge theory \cite{Oe:qlgt}.

The main advantage of the diagrammatic formulation is that it
allows to perform otherwise cumbersome and complicated algebraic
manipulations through simple diagrammatic
identities. This is crucial for our treatment in
Section~\ref{sec:topinv}.
Furthermore, this formulation is manifestly covariant as it does not
require any choice of bases for the representations.

\subsection{Diagrammatic identities}
\label{sec:diaid}

In this section, we introduce the diagrammatic identities that are
instrumental for our new
proof of topological invariance.

\subsubsection{Gauge fixing}
\label{sec:gauge}

In lattice gauge theory, some group integrals in the partition
function can be removed by using the gauge symmetry.
The considered group
variables are set (``gauge fixed'') to an arbitrary element, usually
the unit element.
This applies equally to
discretized BF-theory which
can be considered a weak coupling limit of
lattice gauge theory.
Diagrammatically, gauge fixing of a group variable corresponds
precisely to removing the cable from the corresponding edge.
Thus, the integration over the group element at this edge given by $T$
is replaced by the identity operation corresponding to the action of
the identity element of the group in the relevant representations.

\begin{figure}[h]
\centerline{\hspace{ 0.5cm} \includegraphics[width=7.5cm]{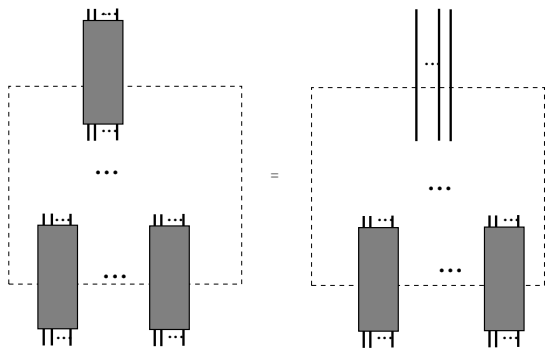}}
\caption{Diagrammatic gauge fixing identity.}
\label{gf2}
\end{figure}

As is well known from lattice gauge theory,
edges can be gauge fixed (i.e., cables removed) precisely as long
as the gauge fixed edges do not form any closed loop.
In fact, this can be seen to follow purely diagrammatically from the
identity depicted in Figure~\ref{gf2}. This identity follows from
Proposition~\ref{prop:T} (see \cite{Oe:qlgt} for details). Note that
gauge fixing is not just an identity of the partition function
$\mathcal{Z}$ as a whole, but of each summand
$\mathcal{Z}_\mathcal{C}$ individually.

\subsubsection{Summation identity}
\label{sec:sumid}

In contrast to the gauge fixing the summation identity
allows the removal not only of a cable but of a wire as
well. It takes the form
\begin{eqnarray}
\sum \limits_{\rho_0} \dim \rho_0 \begin{array}{c}
\includegraphics[width=1.75cm]{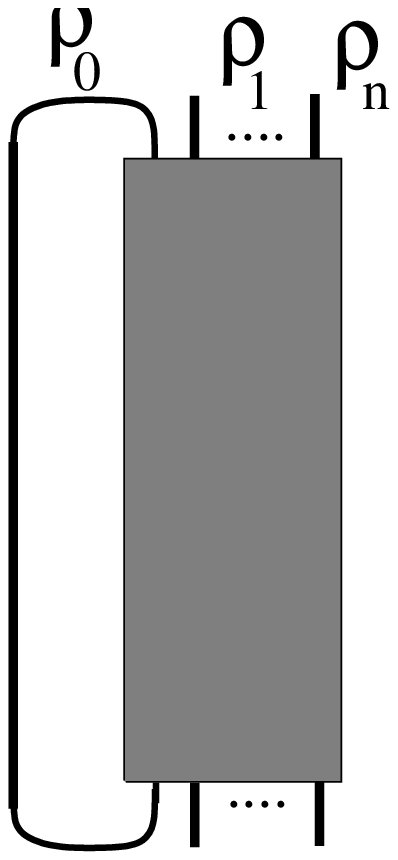}\end{array}=\begin{array}{c}
\includegraphics[width=.99cm]{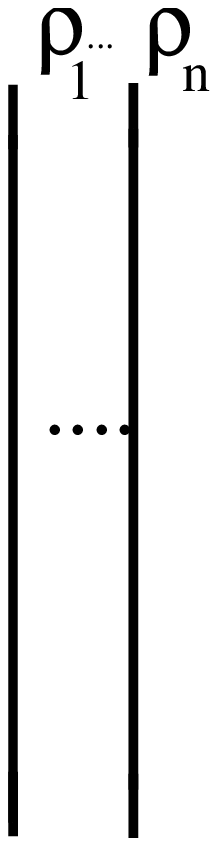}\end{array} .
\label{pito}
\end{eqnarray}
The sum ranges over all finite-dimensional irreducible representations.

In order to prove (\ref{pito})
one uses the fact
that every representation can be decomposed as a direct sum of
irreducible ones (the category of representations of the group is
semisimple). For irreducible representations one can then deduce
(\ref{pito}) 
from the identity
\begin{eqnarray}
\begin{array}{c}
\includegraphics[width=1.2cm]{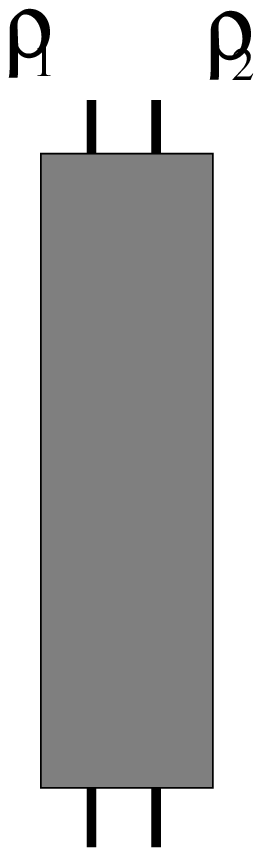}\end{array}=\delta_{{\rho_1}{\rho_2}} \ 
\left({\dim} \rho _{1}\right)^{-1} \begin{array}{c}
\includegraphics[width=.5cm]{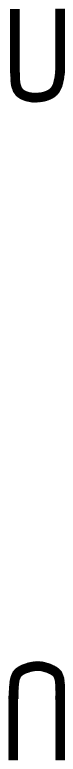}\end{array}.
\end{eqnarray}
See \cite{Oe:qlgt} for details.

%% file: qgrp.tex
\section{Quantum group generalization in 3d}

\label{sec:qg}

Unfortunately, BF-theory as presented in the previous section is
divergent, even in its discretized form. A way to
render the theory finite is to replace groups by $q$-deformed
groups at roots of unity. This can be done in the 3-dimensional case
(to which we confine ourselves in the following) and gives rise to
interesting topological invariants. This was first carried out by
Turaev and Viro using $U_q(\mathfrak{sl}_2)$ \cite{TuVi:inv3}. It was
generalized to the case of spherical quantum groups (or categories) by
Barrett and 
Westbury \cite{BaWe:invplm}. We consider here the (sufficiently general)
case of ribbon quantum groups (or categories).

Crucially, the formalism introduced in the previous section also
permits this generalization.
We only give a rough sketch
here while referring the reader to \cite{Oe:qlgt}
for an extensive treatment with full proofs.

The quantum group case is most conveniently described in a categorial
language. In the group case the relevant category is the category of
finite-dimensional representations of the group. Instead of
starting with a group and then considering its representations we
need to think now of abstractly given objects with
certain operations and properties. More precisely, we use the notion of
\emph{semisimple ribbon category} (in the sense of Turaev
\cite{Tur:qinv}). As each $q$-deformed enveloping algebra
$U_q(\mathfrak{g})$ gives rise to such a category
(after ``purification'' in the root of unity case, see \cite{Tur:qinv})
this is sufficient for our purposes. Note however, that this notion is
more general than that of modular category.

The diagrammatics is modified as follows:
Previously, crossings of lines were simply given by the intertwiner
$V\tens W\to W\tens V: v\tens w\mapsto w\tens v$. Now,
over- and under-crossings are distinct morphisms
(=``intertwiners'') $\psi_{V,W},\psi^{-1}_{V,W}:V\tens W\to W\tens V$,
called the
\emph{braiding}.
Furthermore, lines are framed, i.e., replaced by ribbons. The twist
of a ribbon is a new diagrammatic element corresponding algebraically
to a family of morphisms $\nu_V:V\to V$.
However, in all of the
following we can omit the framings in the diagrams and
implicitly use \emph{blackboard framing}. That means that each line
represents a ribbon which lies face-up and has no twist.

The family of morphisms $T$ is now defined as follows.
Consider the identity morphism $\id_V$ of an object $V$ to itself. It
has a decomposition as a finite sum
\begin{equation}
\id_V=\sum_i f_i\circ g_i
\label{eq:domination}
\end{equation}
where
$g_i:V\to V_i$ and $f_i: V_i\to V$ and $V_i$ are simple objects in the
category (axiom of domination \cite{Tur:qinv}). Now $T_V:V\to V$ is
defined as the
restricted sum $\sum_i'
f_i\circ g_i$ where we only sum over the indices $i$ such that $V_i$
is isomorphic to the trivial object $\one\cong\mathbb{C}$.
The morphisms $T$ have now precisely the properties listed in
Proposition~\ref{prop:T}. Furthermore, they have the trivial braiding
and trivial twisting property
\begin{pro}[\cite{Oe:qlgt}]
\label{prop:qT}
\begin{itemize}
\item[(g)]
 $\psi_{V,W}\circ (T_V\otimes\ 1_W) = \psi^{-1}_{W,V}\circ
 (T_V\otimes1_W)$.
\item[(h)] $\nu_V\circ T_V = T_V = T_V\circ\nu_V$.
\end{itemize}
\end{pro}

Let us now briefly discuss how the partition function is generalized
to this categorial setting. The idea is to use its diagrammatic
representation in terms of the circuit diagram in order to define its
value. However, this is less straightforward now than in the group
case. 
Very roughly, it is achieved by cutting the circuit diagram into pieces
by cutting through each cable, then projecting each piece onto the
plane and
reconnecting the pieces with $T$-diagrams.
The emerging diagram resembles the circuit diagram in a particular
2-dimensional projection.
(See \cite{Oe:qlgt} for details.)
Indeed, as we can think of the identities
used in the following sections as identities between the corresponding
projections, the results derived there apply to the general quantum group
case.

The sum over labellings with irreducible
representations in (\ref{Z1}) becomes a sum over labellings with
equivalence classes of simple objects in the category. At the same
time the
dimensions of the representations are replaced by the quantum
dimensions of the simple objects, given diagrammatically by a loop
labelled by the object. If the category has only 
finitely many inequivalent simple objects the partition function is
manifestly finite. In particular, this is true in the relevant
case of $q$-deformed enveloping algebras
$U_q(\mathfrak{g})$ at roots of unity giving rise to modular
categories.

Importantly, the gauge fixing identity (Section~\ref{sec:gauge}) as
well as the summation identity (Section~\ref{sec:sumid}) generalize
to the quantum group case.
For the gauge fixing this is a consequence of Propositions~\ref{prop:T}
and \ref{prop:qT}. In the summation identity, the dimension
is replaced by the quantum dimension (the loop diagram) and it is proven
using the domination property (\ref{eq:domination}) instead of
conventional semisimplicity.

Although for simplicity we use solely the language of the group
context, the results of the following section apply to the
quantum group context of semisimple ribbon categories in the above
described way.
(In fact, they even apply to semisimple spherical categories, but we do
not consider this here).

%% file: topinv.tex
\section{Generalized proof of topological invariance in 3d}

\label{sec:topinv}

In this section we show, in the 3-dimensional case, that the partition
function of BF-theory is independent
of the chosen cellular decomposition of the manifold $M$.
To be more precise, the invariant quantity corresponds to a rescaling of the 
partition function (\ref{Z1}), namely
\begin{equation}\label{moco}
{\Gamma}=\tau^{-n^{ (0)}}{\zz},
\end{equation}
where $n^{ (0)}$ is the number of 0-cells in $\mathcal{K}$ (corresponding
to the number of 3-cells in
$\mathcal{K}^{*}$) and
$\tau\defeq\sum_{\rho} (\dim\rho)^2$ which is divergent in the Lie
group case.
The rescaling factor is associated to the
non-compact gauge symmetry $B \rightarrow B+d_{ A} \eta$ of BF-theory 
\cite{Fre:fac}.
Gauge equivalent configurations are summed over in (\ref{Z1}). Gauge volume 
factors
should be divided out of the partition function according to 
the standard Faddeev-Popov procedure.

In Section~\ref{sec:dia} we introduced a formulation of quantum BF-theory
for an arbitrary cellular decomposition of $M$. Traditionally, one
restricts to simplicial decompositions $\Delta$ of $M$.
This usually simplifies the proof of topological invariance of (\ref{Z1})
due to the particular combinatorial properties of $\Delta^{*}$: in this case
all cables have only three wires and vertices (0-cells) are four-valent.
Equivalent simplicial complexes are related by a finite set of moves
(Pachner moves). In \cite{Tur:qinv,BaWe:invplm}, $6j$-symbols for
$U_q(\mathfrak{sl}_2)$ as well as  their
main  property (Biedenharn-Elliot), were generalized to any modular
(or spherical)
category; then the invariance of the amplitude was shown by
translating the Pachner moves in terms of transformations of
$6j$-symbols. Generalizing this proof to
arbitrary cellular 
decompositions $\mathcal{K}$ of $M$, would amount to knowing an
infinite number of identities of this kind relating arbitrary
`$nj$-symbols'. In contrast,
our diagrammatic formulation allows for this generalization
by utilizing
the simple gauge fixing and summation identities of the
Section~\ref{sec:diaid}. In turn, it should provide a way of inferring
such `higher' order identities for the appearing `$nj$-symbols'.

In this section we construct the generalized invariance proof by first
concentrating on the special case of simplicial decompositions and then
extending our construction to arbitrary cellular decompositions.

\subsection{The simplicial case}

Topologically equivalent triangulations are related by Pachner moves
\cite{Pac:plmfd}. For showing that (\ref{moco}) is a topological
invariant for simplicial decompositions $\Delta$ it is thus sufficient
to show invariance under Pachner moves.
In order to do this we first need to translate Pachner
moves into the dual complex ${\Delta^{*}}$.

\begin{figure}[h]
\centerline{\hspace{0.5cm}
\(\begin{array}{c}
\includegraphics[width=3cm]{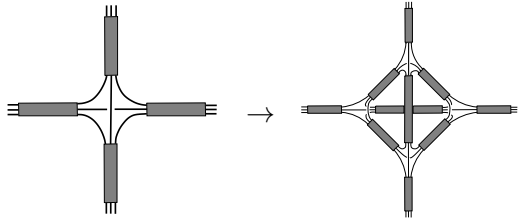}
\end{array} \rightarrow
\begin{array}{c}
\includegraphics[width=3cm]{figi/PM3d31a.eps}
\end{array}\) }
\caption{The (1,4) Pachner move represented by circuit diagrams.}
\label{fig:0}
\end{figure}

In three dimensions there are four Pachner moves: the (1,4) Pachner
move, the (2,3) Pachner move, and their inverses.
The (1,4) move creates four tetrahedra out of one in the following way:
put a point $p$ in the interior of the tetrahedron whose vertices are
labeled $p_i, \ i=1,\dots 4$, add the four $1$-simplices $(p,p_{i})$,
the six triangles $(p,p_{i},p_{j})$ (where $i \ne j$), and the four 
tetrahedra $(p,p_{i},p_{j},p_{k})$
(where $i \ne j \ne k \ne i$). In the dual complex ${\Delta}^{*}$, the 
original
tetrahedron corresponds to a single vertex at which four edges and six faces 
meet.
After the move is implemented, one has four vertices in
${\Delta}^{*}$ (corresponding to the four tetrahedra in $\Delta$)
connected by edges and surfaces to form a tetrahedron
in ${\Delta}^{*}$. In terms of circuit diagrams the move is
illustrated in Fig.~{\ref{fig:0}}.

\begin{figure}[h!]
\centerline{\hspace{0.5cm}
\(\begin{array}{c}
\includegraphics[width=2.2cm]{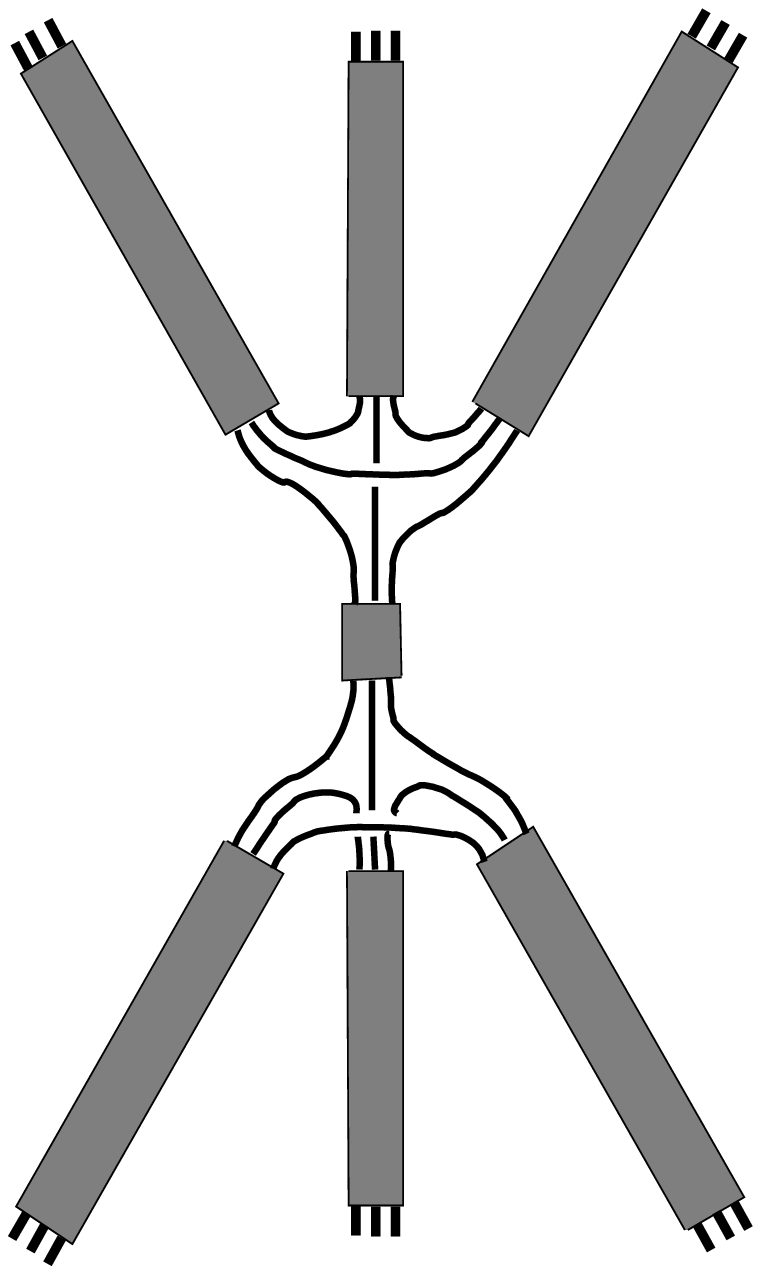}
\end{array} \rightarrow
\begin{array}{c}
\includegraphics[width=4cm]{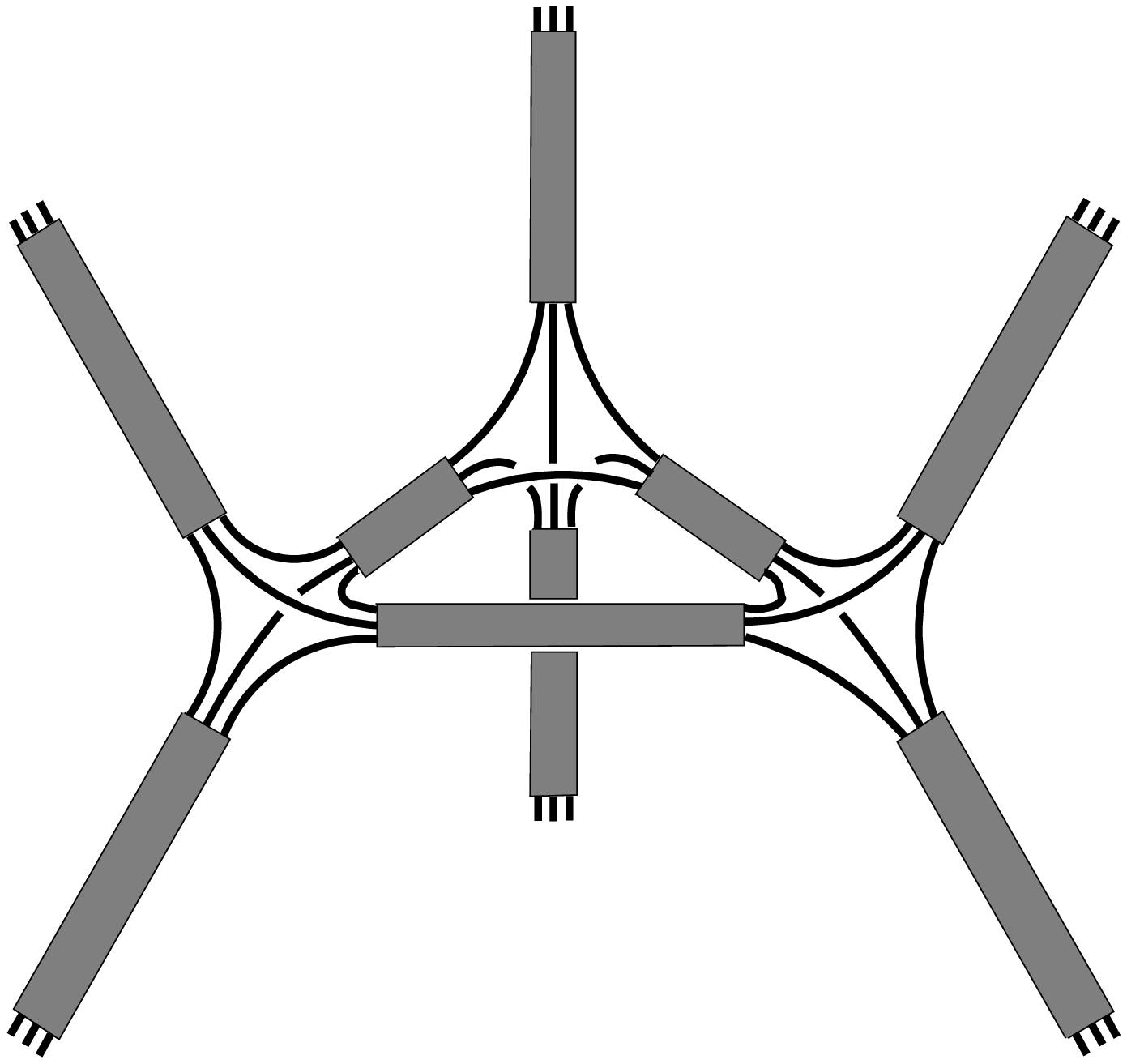}
\end{array}\) }
\caption{The (2,3) Pachner move represented by circuit diagrams.}
\label{fig:00}
\end{figure}

The (2,3) Pachner move consists of the splitting of two
tetrahedra into three tetrahedra. One replaces two tetrahedra 
$(u,p_1,p_2,p_3)$, and
$(d,p_1,p_2,p_3)$, sharing the triangle $(p_1,p_2,p_3)$ with the
three tetrahedra $(u,d,p_{i},p_{j})$ where $i\ne j =1,\ldots 3$.
In terms of circuit diagrams in ${\Delta}^{*}$ the move corresponds to
the one illustrated in Fig.~\ref{fig:00}.

The proof of topological invariance consists of showing that the 
Figs.~\ref{fig:0} and \ref{fig:00}
give rise to identical values of $\Gamma$ in Eq.~(\ref{moco}), as they are 
essentially diagrammatic identities.

\begin{figure}[h]
\centerline{\hspace{0.5cm}
\(\begin{array}{c}
\includegraphics[width=3.5cm]{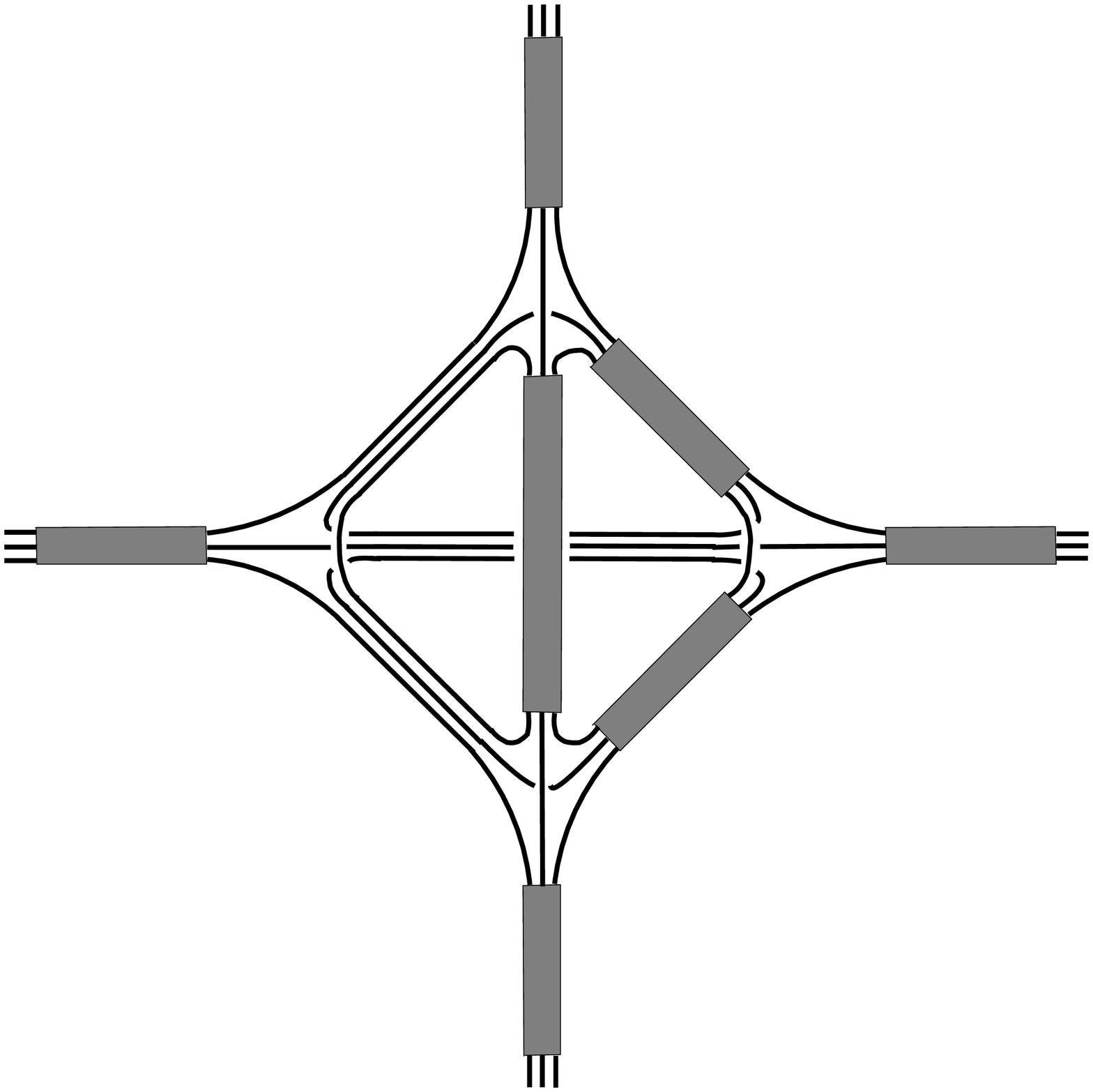}
\end{array} \rightarrow
\begin{array}{c}
\includegraphics[width=3.5cm]{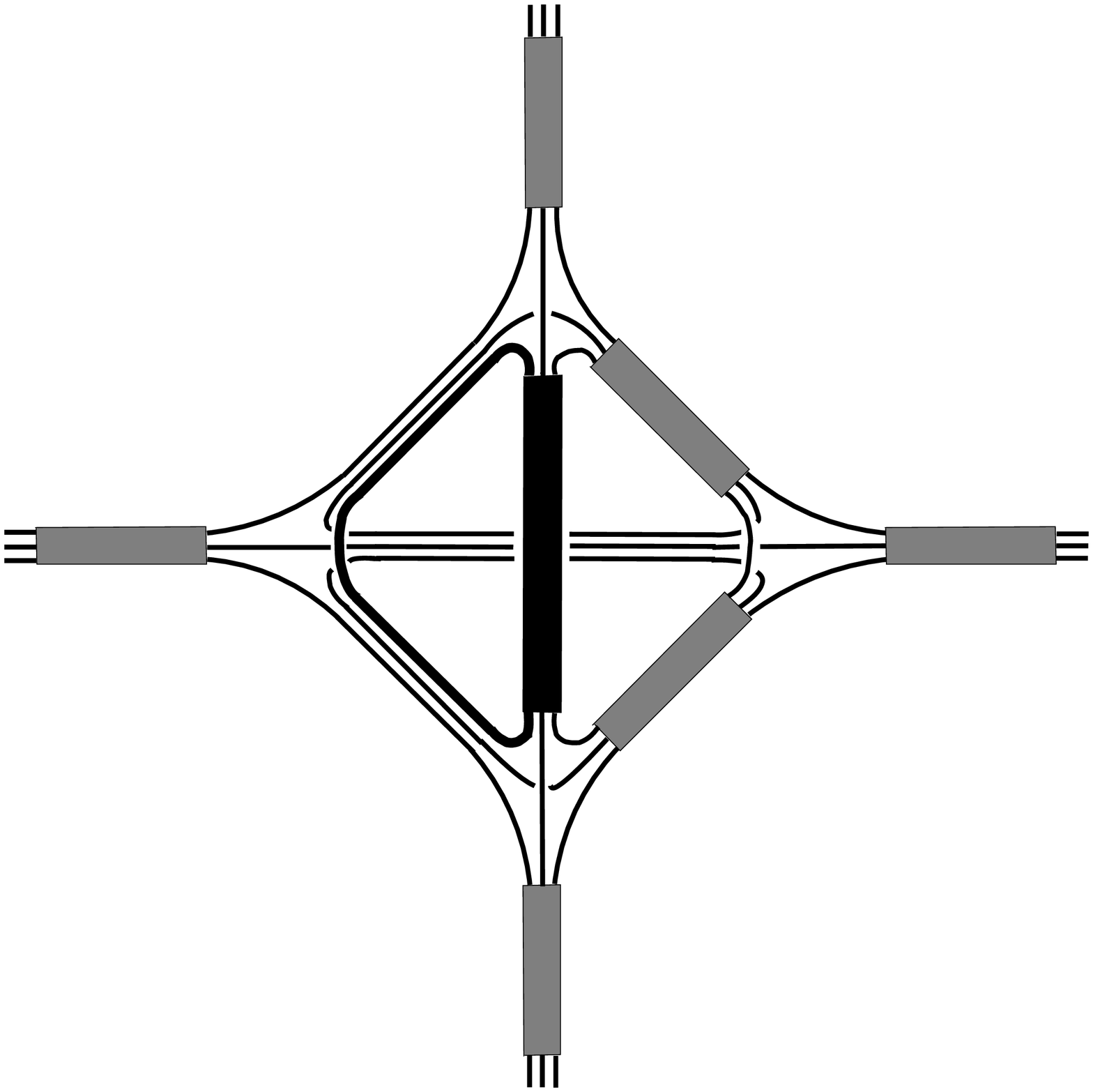}
\end{array}\rightarrow
\begin{array}{c}
\includegraphics[width=3.5cm]{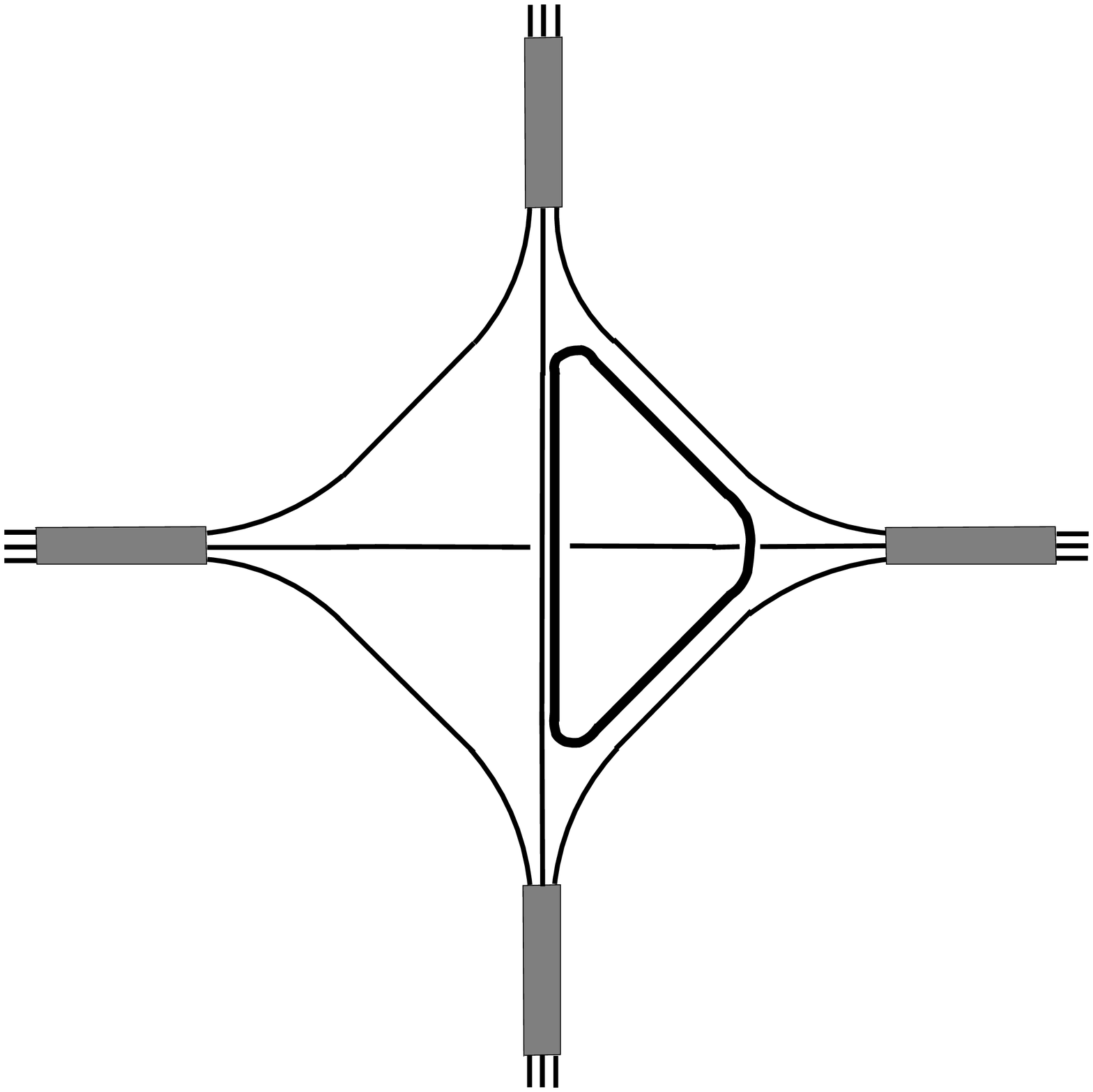}
\end{array}\) }
\caption{Invariance of the partition function under the (1,4) Pachner move.}
\label{fig:1}
\end{figure}

Let us first consider the (1,4) move. We start with the circuit diagram
on the right hand side of Fig.~\ref{fig:0} and we eliminate three
redundant cables using the
gauge fixing. In this way we obtain the first diagram on the left hand
side of
Fig.~{\ref{fig:1}}.
We then use the summation identity (\ref{pito}) to erase all the remaining
internal cables as follows. The factor ${\rm dim} \rho$ in (\ref{pito}) and 
the corresponding
sum over representations is provided by the partition function (\ref{Z1}).
In the second diagram in Fig.~{\ref{fig:1}} one such
internal cable and corresponding closed wire are emphasized.
After the three internal cables are eliminated
in this way, we obtain the circuit diagram on the right hand side of
Fig.~{\ref{fig:1}}.
This diagram corresponds precisely to the one tetrahedron diagram on the
left of Fig.~\ref{fig:0} multiplied by a closed loop evaluation.
Such a closed loop corresponds to the trace of the identity in the
corresponding representation, i.e., ${\rm dim} \rho$,
which together with the factor ${\rm dim} \rho$ in (\ref{Z1})
gives an overall factor $\tau=\sum_{\rho} (\dim\rho)^2$.
In the process the number $n^{(0)}$ of 3-cells in ${\Delta}^{*}$
(0-cells in $\Delta$) has 
decreased by one;
therefore, this extra $\tau$ precisely compensates
for this change and (\ref{moco}) remains invariant.

\begin{figure}[h]
\centerline{\hspace{0.5cm}
\( \begin{array}{c}
\includegraphics[width=3cm]{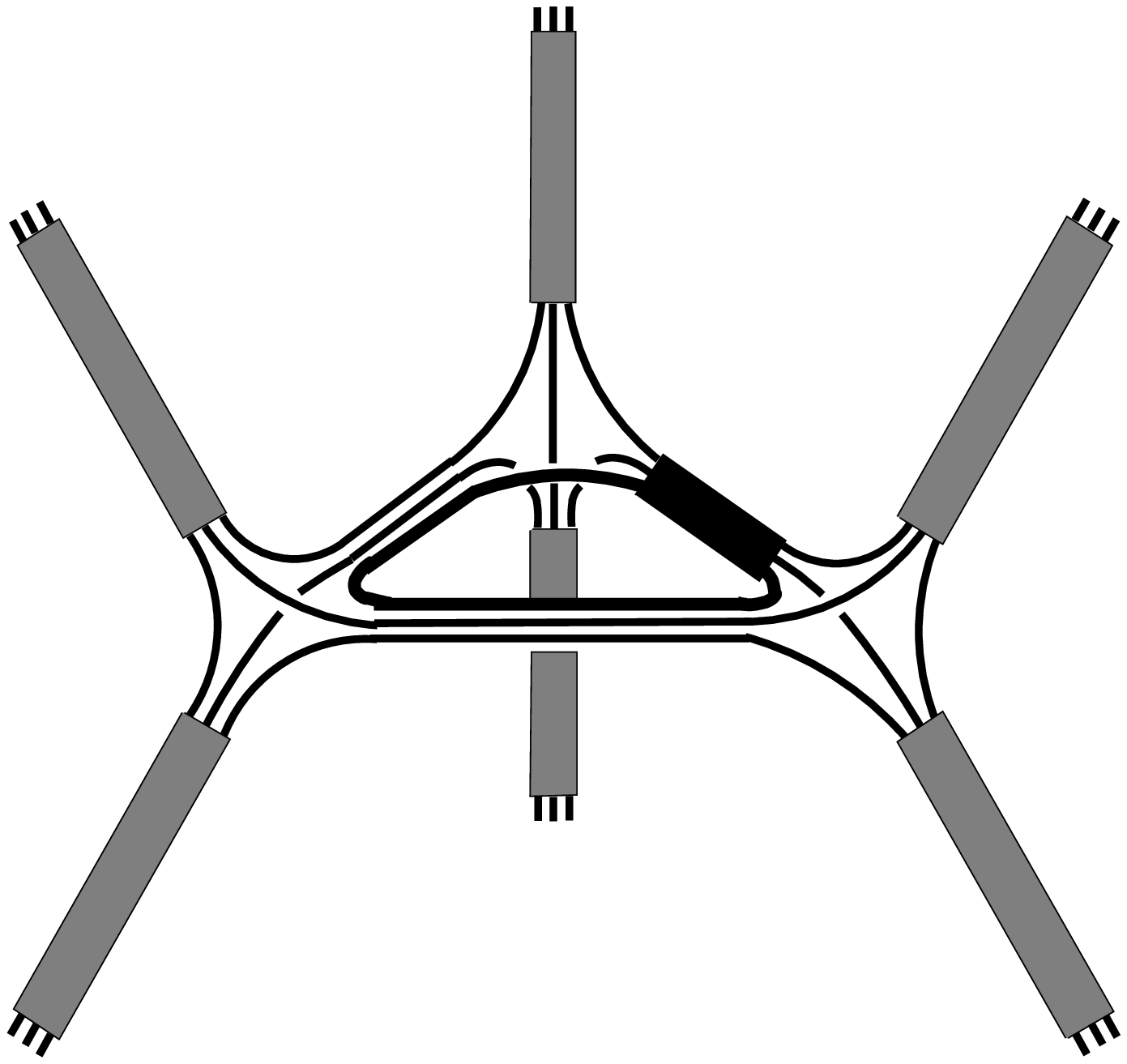}
\end{array} {\bf \rightarrow}
\begin{array}{c}
\includegraphics[width=3cm]{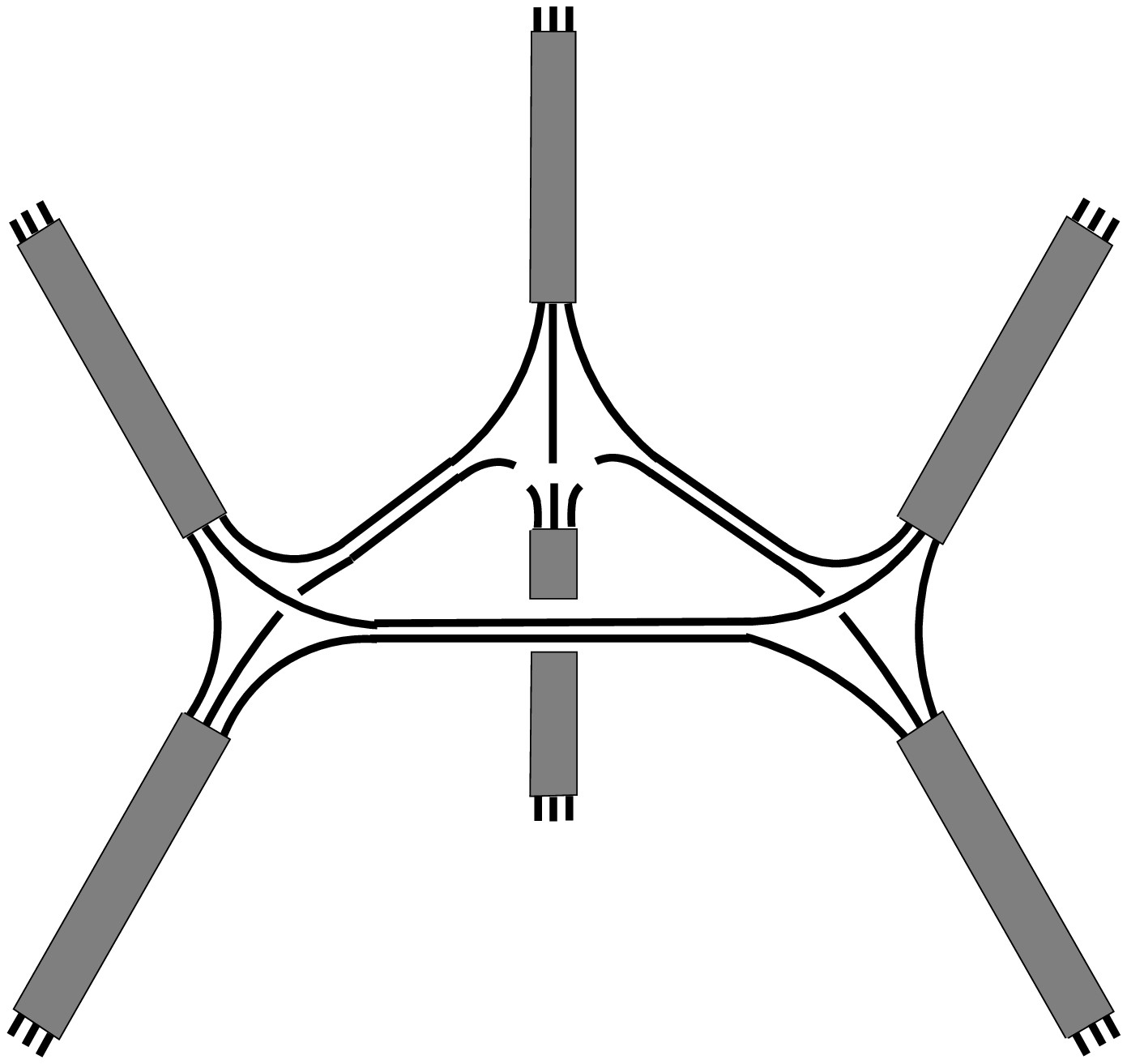}
\end{array}
{\bf \rightarrow}
\begin{array}{c}
\includegraphics[width=1.8cm]{figi/PM3d32d.eps}
\end{array}\)}
\caption{Invariance of the partition function under the (2,3) Pachner move.}
\label{fig:5}
\end{figure}

Using the same techniques we prove invariance of (\ref{moco})
under the (2,3) Pachner move. We start with the diagram on the right
hand side of 
Fig.~\ref{fig:00} and eliminate two of the
internal cables by means of gauge fixing. In this way we obtain the
left hand side
diagram on Fig.~{\ref{fig:5}},
where we have emphasized the remaining internal cable and its corresponding 
closed wire.
The latter can be eliminated using (\ref{pito}), see the diagram in
the center of
Fig.~{\ref{fig:5}}. In the diagram on the right hand side of
Fig.~{\ref{fig:5}} we have deformed the previous diagram and
we have added a redundant cable (by inverse gauge fixing) to obtain
the circuit diagram corresponding to
the diagram on the left hand side of Fig.~{\ref{fig:00}}.
The number of 3-cells does not change under this move.

We have proven the following result:
\begin{pro}
\label{prop:invs}
Consider two simplicial decompositions $\Delta$
and $\Delta^{\prime}$ of a 3-dimensional manifold $M$. 
Then the following equality holds:
\[
\Gamma(\Delta)=\Gamma(\Delta^\prime),
\]
where the value of the partition function is computed according to 
(\ref{moco}) on the
corresponding complex.
\end{pro}

%% file: geninv.tex
\subsection{General cellular decompositions}

\label{sec:geninv}

We now generalize the above results from simplicial to general cellular
decompositions of a 3-dimensional manifold.
We make use of the following result (see \cite{rour}).
\begin{pro}
\label{prop:1}
A cellular complex $\cal K$ can be subdivided into a simplicial complex
$\Delta_{\cal K}$ without introducing any new vertices (0-cells).
\end{pro}

We start by proving the following lemma.
\begin{lem}
\label{lem:dec}
Let $\cal K$ be a cellular decomposition of a 3-dimensional manifold
$M$ and $\Delta_{\cal K}$ an associated
simplicial decomposition (Prop.~\ref{prop:1}).
Then, the following equality holds
\[{\Gamma}({\cal K})={\Gamma}(\Delta_{\cal K}).\]
\end{lem}
Note that even ${\cal Z}({\cal K})={\cal Z}(\Delta_{\cal K})$ 
as $n^{(0)}$ is the same
for ${\cal K}$ and $\Delta_{\cal K}$.

This lemma is proven using a finite set of moves relating a
cellular decomposition with an associated simplicial decomposition.
For definiteness we consider these moves as leading from the
simplicial subdecomposition to the cellular decomposition.

\begin{figure}[h]
\centerline{\hspace{0.5cm}
\(\begin{array}{c}
\includegraphics[width=4.3cm]{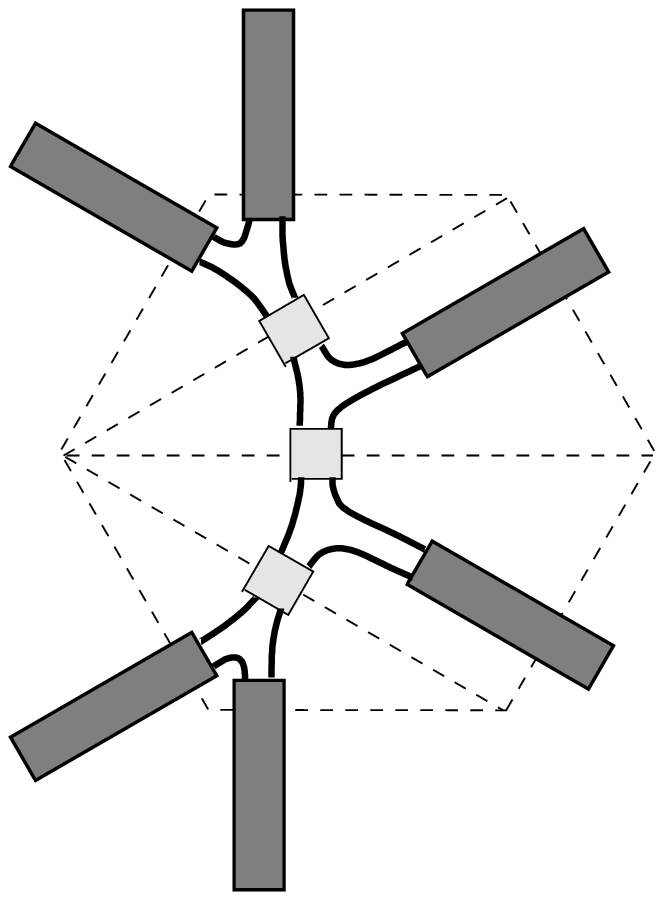}
\end{array} \rightarrow
\begin{array}{c}
\includegraphics[width=3cm]{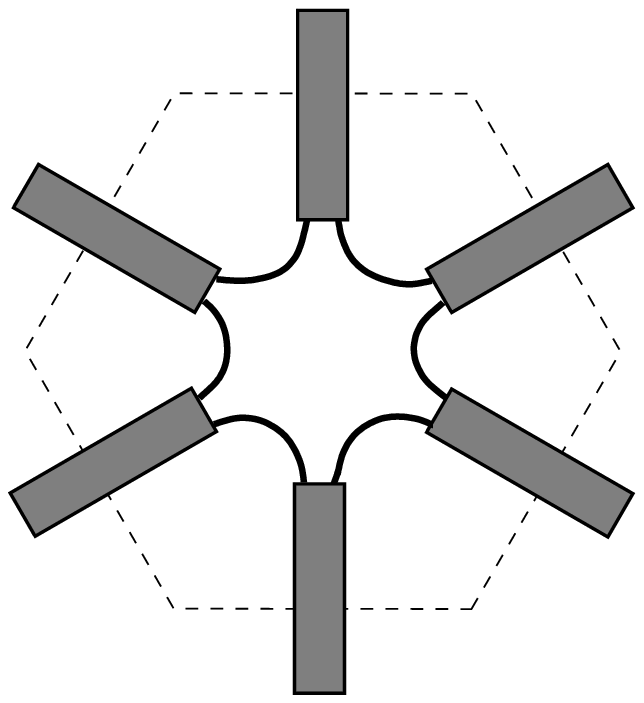}
\end{array} \) }
\caption{Lemma \ref{lem:dec} in two dimensions.}
\label{exa}
\end{figure}

For illustration let us consider first the 2-dimensional case.
There is just one move, namely forming the union of two 2-cells by
removing a common boundary (the inverse consisting of the
subdivision of a 2-cell). That this move leaves the partition function
invariant is a straightforward consequence of the gauge fixing identity.
See Fig.~\ref{exa} for illustration.

The 3-dimensional case of interest to us here is more involved. The
relevant moves are described by the following proposition.

\begin{pro}
\label{prop:mov}
Let $\mathcal{K}$ be a cellular decomposition of a 3-manifold and
$\Delta_\mathcal{K}$ an associated simplicial decomposition. Then,
$\mathcal{K}$ and $\Delta_\mathcal{K}$ are related by a sequence of
moves and their inverses from the following list:
\begin{description}
\item[3-cell fusion] Let $\tau$ be a 2-cell which bounds two distinct
3-cells $\sigma$, $\sigma'$. Remove $\tau$, $\sigma$ and
$\sigma'$ and insert the new 3-cell $\sigma''\defeq
\sigma\cup\tau\cup\sigma'$.
\item[2-cell retraction] Let $\mu$ be a 1-cell that bounds only the
2-cell $\tau$ which in turn bounds only the 3-cell $\sigma$. Remove
$\mu$, $\tau$, $\sigma$ and insert the new 3-cell $\sigma'\defeq
\sigma\cup\tau\cup\mu$.
\end{description}
\end{pro}
\begin{proof}
We show that starting with the simplicial subdecomposition we can
recover the original cellular decomposition by applying the stated
moves. We proceed in two main steps: First we show that we can remove
all the subdivisions in all 3-cells of the original
decomposition. Second, we show that we can remove the subdivisions
on the boundaries of the 3-cells.

Thus, consider a 3-cell $\sigma$ together with its subdivision
$\Delta_\sigma$. Assume that
$\Delta_\sigma$ has internal 2-cells. (Otherwise it would consist only
of one 3-cell and we are done.) We claim that we can always remove one
such 2-cell by
either applying the 3-cell fusion move or the 2-cell retraction
move. It is thus sufficient to show that if we cannot apply the
3-cell fusion move, we can apply the 2-cell retraction move.
The only situation that prevents us from applying the 3-cell fusion
move is when every internal 2-cell is bounded on both sides by the same
3-cell. Assume this.
Consider the union of all internal 2-cells and their bounding
1-cells. This forms a branched surface. Consider its boundary. If it
was contained in the boundary of $\sigma$ it would divide $\sigma$ into
at least two disjoint 3-cells contrary to the assumption. Thus, the
boundary of the branched surface must contain 1-cells internal to
$\sigma$. Pick
such a 1-cell. By construction this bounds only one (internal)
2-cell. Thus, we can apply the 2-cell retraction move to it. This
completes step one of the proof.

For the second step we introduce an auxiliary move as follows.
Let $\mu$ be a 1-cell that bounds only the
distinct 2-cells $\tau$ and $\tau'$. Remove $\mu$, $\tau$, $\tau'$ and
insert the new 2-cell $\tau''\defeq \tau\cup\mu\cup\tau'$. We call
this the \emph{2-cell fusion} move. It can be obtained as a sequence
of 3-cell fusion and 2-cell retraction moves as follows.
Consider the (not necessarily distinct) 3-cells $\sigma,\sigma'$ that
are bounded by $\tau$. Indeed they are also bounded by $\tau'$ as
$\mu$ does not bound other 2-cells than $\tau$ and $\tau'$. Consider
the boundary of $\tau\cup\mu\cup\tau'$ and introduce a 2-cell $\tau''$
with this boundary subdividing $\sigma$ into two 3-cells $\sigma''$
(bounded by the 2-cells $\tau,\tau',\tau''$) and (using the
previous name) $\sigma$ by the inverse 3-cell fusion move. By
construction, $\sigma'$ and $\sigma''$ are distinct. Thus, we can
remove $\tau$ via a 3-cell fusion move, denoting the resulting
3-cell again by $\sigma'$. As $\mu$ now only bounds $\tau'$ which in turn
only bounds $\sigma'$ we can apply the 2-cell retraction move to
eliminate $\mu$ and $\tau'$ to arrive at the desired configuration.

Now consider a 2-cell $\tau$ in the boundary of a 3-cell of the
original cellular complex. As the subdivision of $\tau$ does not contain
any internal vertices every internal 1-cell bounds two distinct
2-cells. We can thus apply the 2-cell fusion move to remove all
internal 1-cells one by one. This completes the proof.
\end{proof}

\begin{figure}
\centerline{\hspace{0.5cm}
\(\begin{array}{c}
\includegraphics[width=4cm]{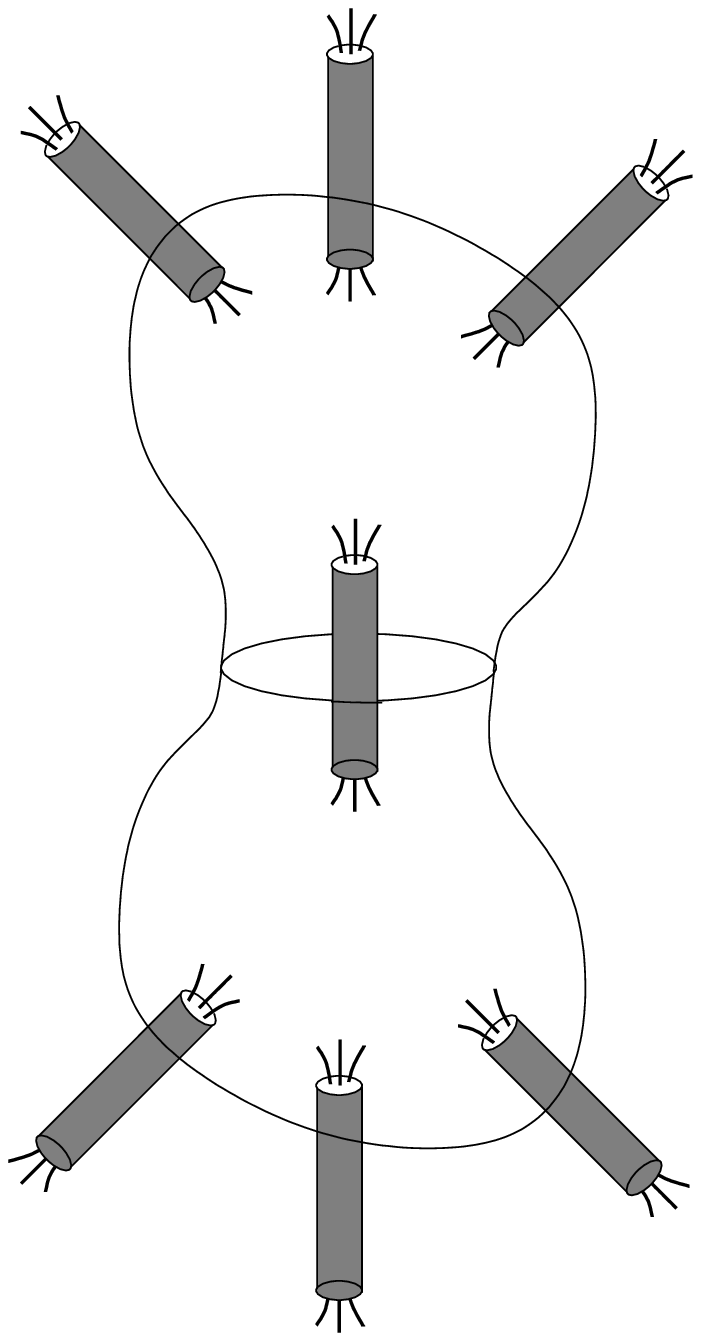}
\end{array} \rightarrow
\begin{array}{c}
\includegraphics[width=4cm]{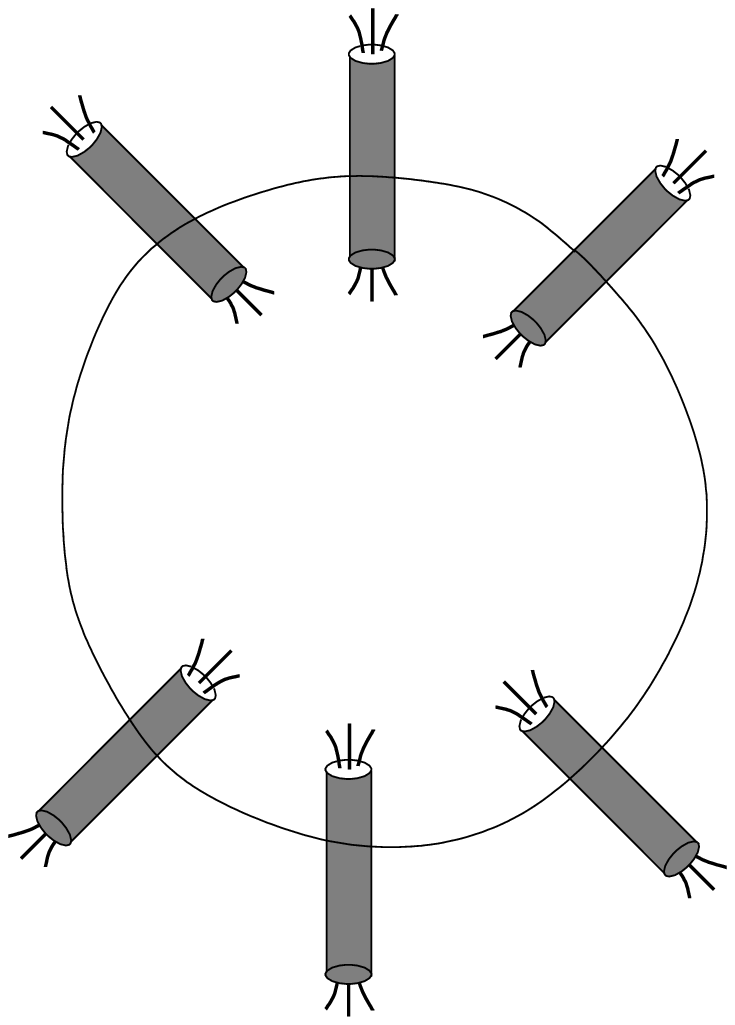}
\end{array} \) }
\caption{The 3-cell fusion move.}
\label{fig:3fuse}
\end{figure}

\begin{figure}
\centerline{\hspace{0.5cm}
\(\begin{array}{c}
\includegraphics[width=4cm]{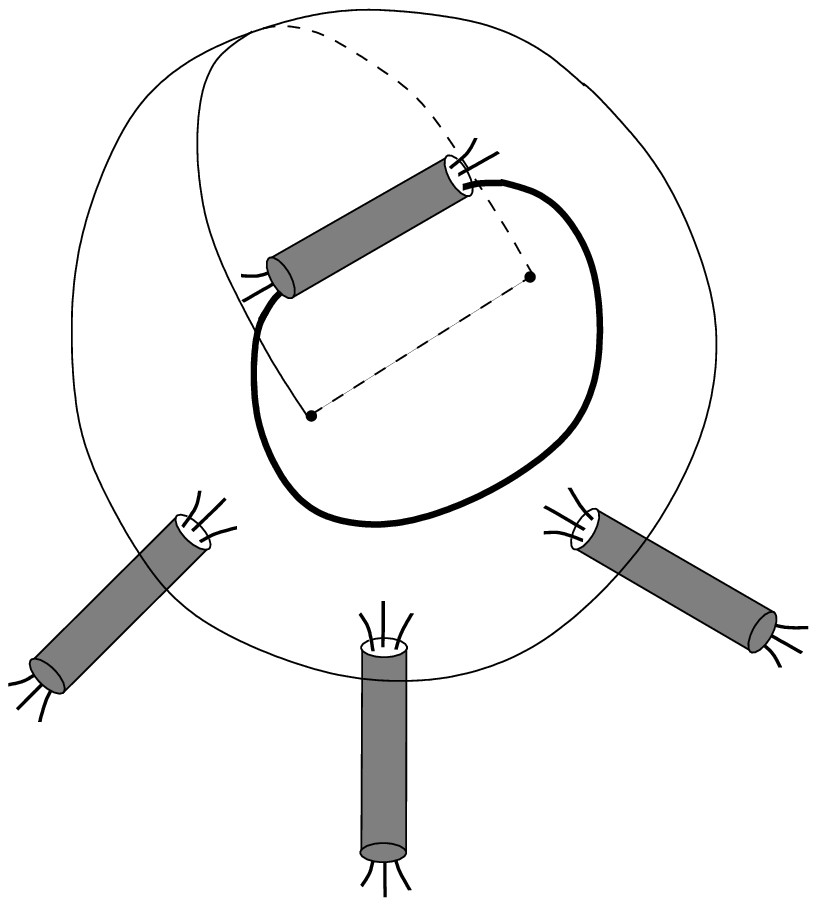}
\end{array} \rightarrow
\begin{array}{c}
\includegraphics[width=4cm]{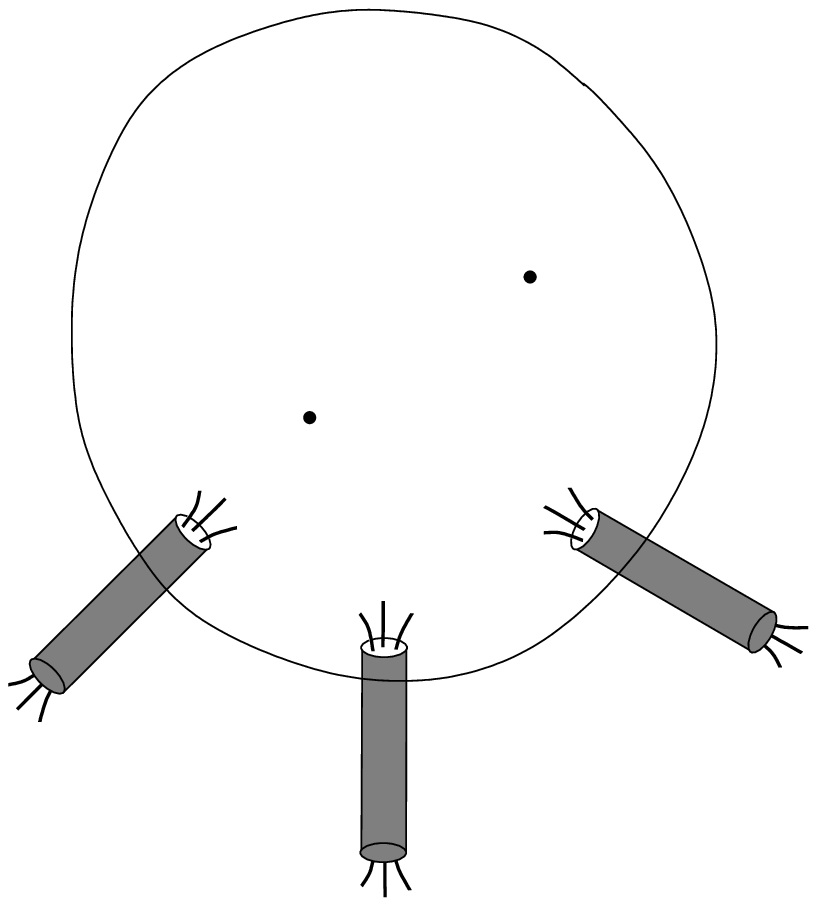}
\end{array} \) }
\caption{The 2-cell retraction move.}
\label{fig:2retract}
\end{figure}

\begin{proof}[Proof of Lemma~\ref{lem:dec}]
It is now sufficient to show that the moves of
Proposition~\ref{prop:mov} leave the partition function
invariant. Consider the Figures~\ref{fig:3fuse} and
\ref{fig:2retract}.
These show the two moves together with relevant
pieces of the circuit diagram.

The 3-cell fusion move is depicted in
Figure~\ref{fig:3fuse}. The 2-cell that separates the initial 3-cells
carries one cable. The removal of the cable corresponds precisely to
the move. It is possible due to the
gauge fixing identity (Figure~\ref{gf2}) and leaves not only
$\mathcal{Z}$ but even $\mathcal{Z}_\mathcal{C}$ invariant.
The dotted line in
Figure~\ref{gf2} corresponds to the boundary of one of the initial
3-cells in Figure~\ref{fig:3fuse}.

The 2-cell retraction move is depicted in
Figure~\ref{fig:2retract}. Diagrammatically it corresponds to removing
the cable that is carried by the 2-cell to be deleted, together with
the wire that goes around the 1-cell to be deleted. This is an
invariance of $\mathcal{Z}$ due to
the summation identity (\ref{pito}). The summation over the
representations in (\ref{pito}) with weight the (quantum) dimension is
precisely contained in the partition function (\ref{Z1}).
\end{proof}

Combining Proposition~\ref{prop:invs} with Lemma~\ref{lem:dec} we
have the following theorem:

\begin{thm}
Given two cellular decompositions ${\cal K}$ and ${\cal K}^{\prime}$ of a
3-dimensional manifold $M$, the following equality holds:
\begin{equation}
{\Gamma}({\cal K})={\Gamma}({\cal K}^{\prime}),
\end{equation}
where the value of the partition function is computed according to 
(\ref{moco}).
\end{thm}            

%% file: conclude.tex
\section*{Concluding Remarks}
In this final section we offer some concluding remarks and consider
possible developments.

Invariance of the partition function under Pachner moves was proven 
originally in terms of the Biedenharn-Elliot identities for $6j$-symbols.
Conversely, one could deduce these identities from the invariance of
the state sum. Our diagrammatic generalization to arbitrary cellular 
decompositions allow us to infer, in this way,  an infinite number of 
relations among `$nj$-symbol'. The precise structure of these
`generalized' Biedenharn-Elliot identities remains to be studied.

\begin{figure}[h]
\centerline{\hspace{0.5cm}
\( \begin{array}{c}\includegraphics[width=3.0cm]{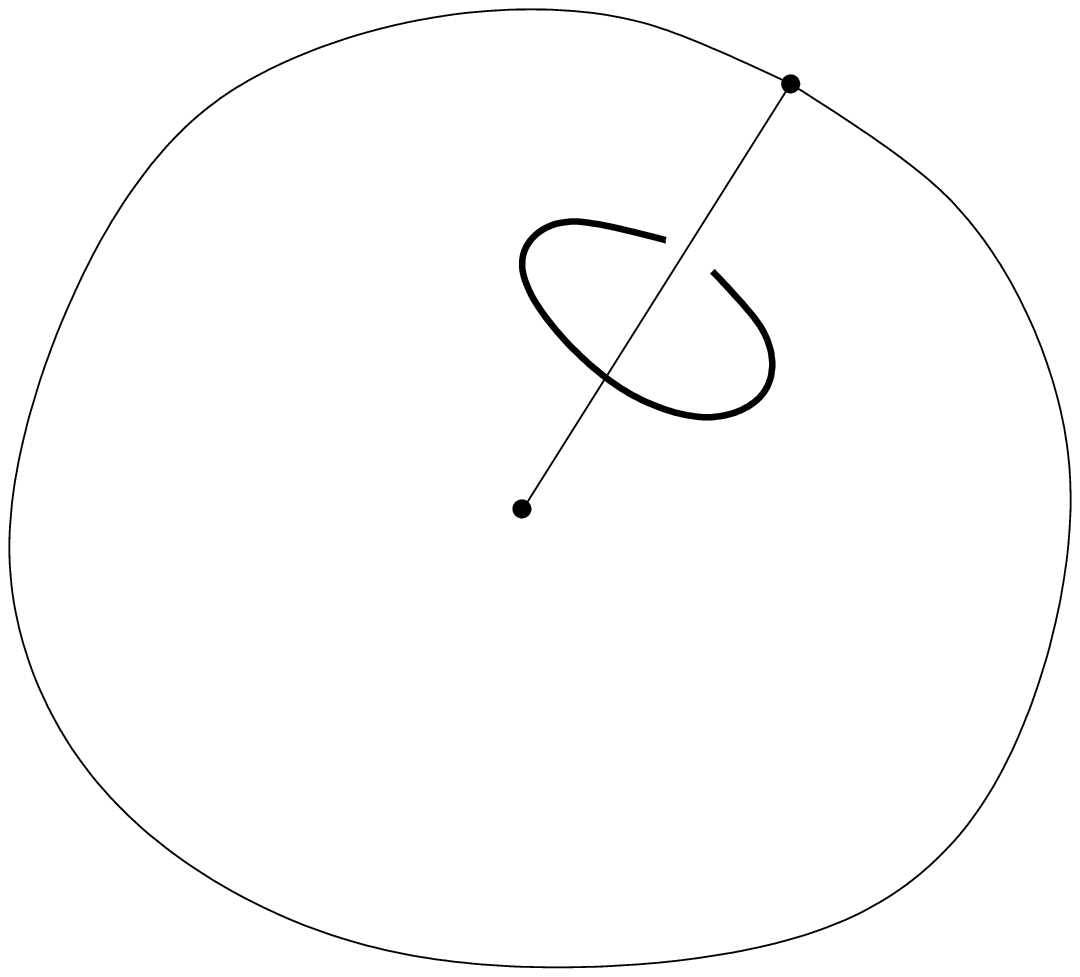}\end{array}
{\bf  \rightarrow}
\begin{array}{c}\includegraphics[width=3.96cm]{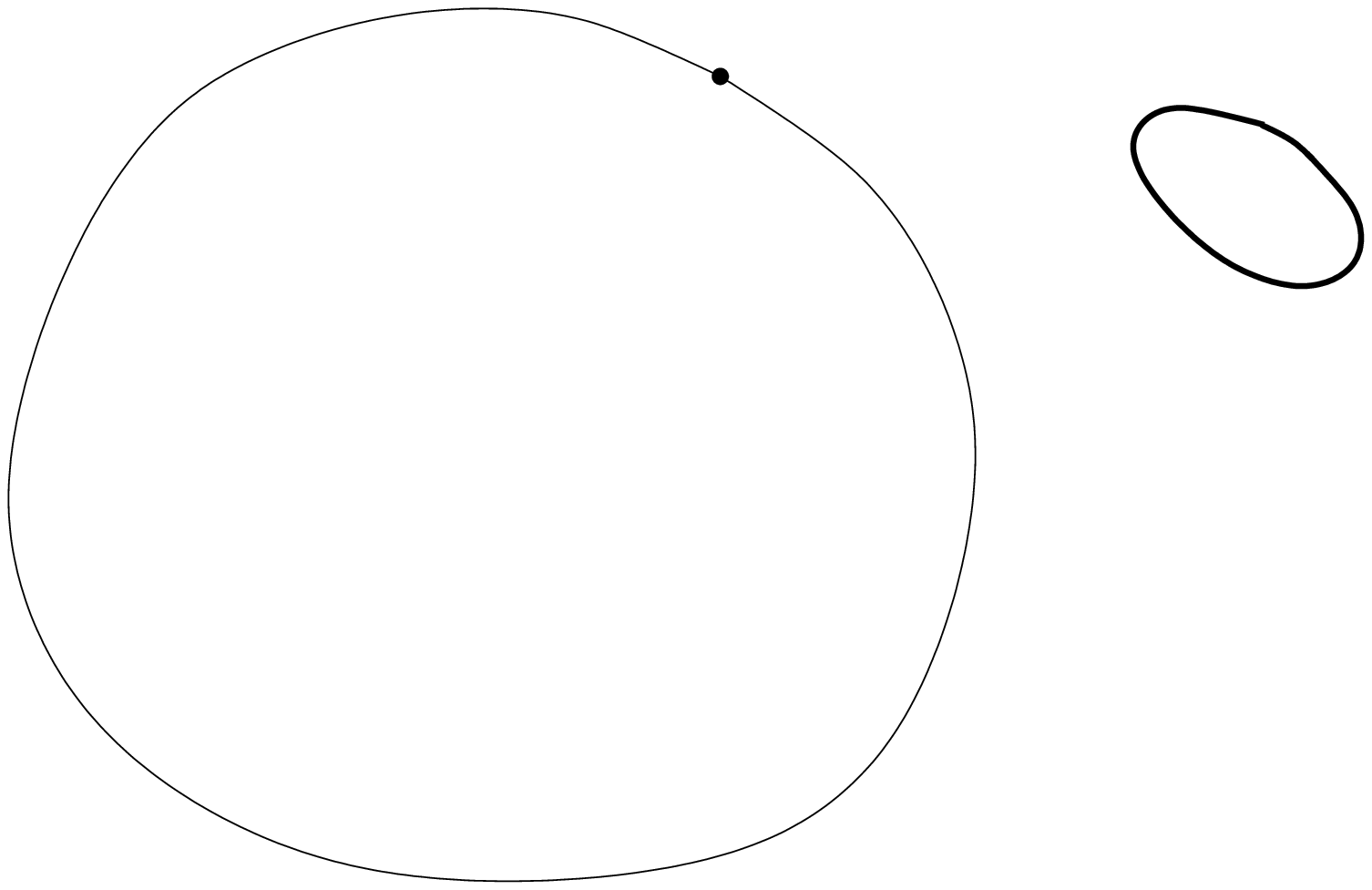} \end{array}\)}
\caption{1-cell retraction move.}
\label{fig:1retract}
\end{figure}

In the proof of topological invariance we have started with simplicial
decompositions (related by Pachner moves) as this is the more familiar
setting. Instead we could have carried out the proof directly with
cellular decompositions alone. To this end we would need a complete
set of moves relating cellular decompositions in 3d. One could wonder
whether the moves given in Proposition~\ref{prop:mov} are already
sufficient for this. Indeed, the (2,3) Pachner move can be obtained as
a sequence of these moves. However, the (1,4) Pachner move cannot.
But it can, if we add just one more move: 
The 1-cell retraction move. This move removes a 0-cell and a 1-cell
inside a 3-cell and is defined analogous to the 2-cell retraction
move. See Figure~\ref{fig:1retract}. In the diagrammatic
picture for the (1,4) Pachner move this is reflected in the presence
of the loop that is extracted from the last tetrahedron (see Figure
\ref{fig:1}). In summary, we have thus shown that any two cellular
decompositions of a compact 3-manifold are related by a sequence of
the following moves: 3-cell fusion, 2-cell retraction, 1-cell
retraction. The generalization of these moves to higher dimensions
should be investigated.

The advantage of cellular decompositions over simplicial ones might
appear a purely technical one in the topological context. However, the
story becomes very different in the non-topological situation, such as
for realistic models of quantum gravity in 4 space-time
dimensions. In particular, the diagrammatic methods considered
here are still applicable for many of the recently proposed models
which are modifications of 4d BF-theory 
\cite{BaCr:relsnet, BFtoGR1,BFtoGR2,BFtoGR3}.
At first sight one might expect that cellular decompositions
(being more general) just make things more complicated.
But exactly the opposite is suggested by our present work. We think
here in particular of the elusive question of renormalization or
``coarse-graining'', see \cite{Mar:alggrain}. In that context,
renormalization identities would
presumably be connected to elementary moves between complexes.
One could imagine these renormalization identities as ``deformations''
of the diagrammatic identities in the topological case. We have
shown that these identities are much simpler for the cellular moves
than for the Pachner moves, suggesting renormalization (rather
surprisingly) to be more manageable in the general cellular case.

When one applies the state sum model of BF-theory to manifolds
with boundaries this naturally defines a TQFT. In the simplicial case,
boundary data are encoded in three-valent spin networks based on the
graphs that 
are dual to the triangulation of the boundaries. As our generalization
allows for general cellular decomposition of the boundaries,
boundary data can be encoded in spin network states of arbitrary
valence \cite{Oe:qlgt}.
A simple application of this generalization is the proof of
equivalence
between different spin network representations of the same physical
state
(due to representation theory skein relations \cite{lectures}). Since
skein relations are equations between spin networks generally involving 
arbitrary graphs, their proof in the state-sum context can only be realized
in terms of our generalization \cite{talk}. The relation between the
state sum (spin foam) approach, the canonical approach and the issue
of the continuum limit
\cite{zapata} might become more transparent in terms of arbitrary
cellular decompositions.

Generalizations of the techniques introduced here to the case of
non-compact groups are of great relevance. Models for Lorentzian quantum
general relativity have been introduced in \cite{Lor1,Lor2}
as constrained state sums of BF-theory on $SL(2,\ccc)$. The models have been
proven to be well defined for a given discretization using the undeformed
group \cite{Lor3}. 
Although the generalization of our identities to infinite dimensional
representations is not obvious, a modification involving ``volume
factors'' should be valid at least for ordinary groups
such as $SL(2,\ccc)$.
The investigation of scaling properties of these
models via our technique might then be possible.

Further issues to be investigated are the application of the
diagrammatics in the context of spin foams generated by field theories
\cite{ReRo:stfeyn} and for defining supersymmetric spin foam
models. See \cite{Oe:qlgt} for more remarks on these points.

%% file: acknowledge.tex
\subsection*{Acknowledgements}

R.~O.\ acknowledges financial support through a NATO fellowship grant.
A.~P.\ acknowledges financial support of the Andrew Mellon Predoctoral 
Fellowship. This work was supported in part by NSF Grants No. 
PHY-9900791 and PHY0090091, and the Eberly research funds of Penn State.